\documentstyle[prb,twocolumn,aps]{revtex}

\begin{document}
\tolerance 10000

\draft

\title{
Electron-Hole Correlation in Fractional Quantum Hall Systems
}
 
\author{P. B\'eran}
\vspace*{0.5truecm}

\address{
Institut de Physique, Universit\'e de Neuch\^atel,
CH-2000 Neuch\^atel, Switzerland
}

\twocolumn[
\date{Received}
\maketitle
\widetext

\vspace*{-1.0truecm}

\begin{abstract}
\begin{center}
\parbox{14cm}{ 
The 2D system of electron confined to the
lowest Landau level is described using a representation of the
density matrix depending both on electron and hole coordinates.
Condensation of the electron system into a fractional quantum Hall state
is associated to clustering of particle and hole coordinates.
The correlation of particle and hole coordinates is studied
and groundstate wave functions are derived for $\nu=p/(2p \pm 1)$.
These wave functions prove to be accurate for the studied cases $\nu=2/3$
and $3/5$ and are identical to {\it pair} wave functions
[R. Morf {\it et al.}, Phys. Rev. {\bf B34}, 3037 (1986)]
for $\nu=2/3$ and $2/5$.
}
\end{center}
\end{abstract}

\pacs{
\hspace{1.9cm}
}
]

\narrowtext

The theory of Laughlin \cite{Laughlin83} provides a good
microscopic description
of the $\nu=1/m$ incompressible quantum Hall states, with $m$ odd.
Based on this description, it has been proposed to characterize
fractional quantum Hall states by the occurence
of binding of flux quanta to electrons \cite{Read89}.
An explicit formulation for such a binding
is provided by the composite Fermion approach \cite{Jain89,Halperin93},
which makes use of a singular gauge transformation
to map the system of electrons in presence of magnetic field onto
a system of composite Fermions, which are objects composed
of an electron bound to an even number of flux quanta.

In this note, an alternative formulation is presented for treating
the binding of flux quanta to electrons.
This approach makes use of a particular representation
of the finite temperature density matrix
for the electron system confined to the lowest Landau level
which consists in matrix elements of the density matrix operator
evaluated between states parametrized by electron coordinates
and states parametrized by the coordinates of electron vacancies,
i.e. holes.
The latter coordinates can be regarded as positions of flux quanta.

This formulation, in which electron-hole symmetry
is fully explicit, leads to a simple and natural picture for the
incompressible quantum Hall state at $\nu=p/q$:
At some finite temperature, the density matrix is large only for
configurations in which electron
and hole coordinates can be grouped into clusters containing each
$p$ electron coordinates and $q-p$ hole coordinates.
This electron-hole binding results dynamically from the repulsion
between electrons.
The creation of a neutral excitation corresponds to the breaking
of such a cluster into two smaller clusters, each one leading to
a local fractional charge defect.
The fact that this breaking requires a finite amount of potential energy
leads to a gap in the energy spectrum and to incompressibility.
The off-diagonal long range order characterizing incompressible
groundstates \cite{Girvin87} results from the fact that
these clusters are Bosons and Bose-condense.

The advantage of working with this finite temperature description
is that the correlation between
hole coordinates, which play the role of positions of flux quanta,
and electron coordinates can be written down explicitely.
However, the price to pay is the work necessary to derive from this
density matrix a microscopic description of the system a zero temperature.
We show how to obtain groundstate wave functions either as a function
of electron coordinates or as a function of hole coordinates
in the case of the fractional quantum Hall states at $\nu=p/(2p+1)$. 
The case is made that the wavefunctions derived from this density matrix
are generalizations of the pair wavefunctions of Ref. \cite{Morf86},
which are known to be accurate.
In the latter wave functions, the pairing of electrons can be regarded
as resulting directly from the presence in the density matrix of clusters
containing each $p=2$ electrons and $q-p$ holes.

This description of fractional quantum Hall states based on a density matrix
is in fact analogous to the one provided by the composite Fermion approach
\cite{Jain89,Halperin93}.
Indeed, a cluster made up of one electron and one hole corresponds
to a composite Fermion and a cluster made up of $p$ electrons and $p+1$
holes, which enters into the description of the $\nu=p/(2p+1)$ state,
corresponds to $p$ composite Fermions, each one belonging to a different
Landau level, all bound to a flux quantum of the field experienced by composite
Fermions.

Let us first study the density matrix of the 2D electron gas
confined to the lowest Landau level
at a filling factor $\nu=p/q$ characterized by the presence of an
incompressible groundstate.
More specifically, we consider the quantity
\begin{equation}
\rho_{\beta}(w_1...w_{hN} z_1...z_{pN}) =
\left< w_1...w_{hN} \right|
e^{ - \beta \tilde{V} }
\left| z_1...z_{pN} \right>
\label{eq:dens_mat_def}
\end{equation}
where $\tilde{V}$ is the electron-electron interaction projected
onto the lowest Landau level and where
$\left| z_1...z_{pN} \right>$ and
$\left| w_1...w_{hN} \right>$ are states respectively
given in electron representation and hole representation, defined by
\begin{eqnarray}
\left| z_1...z_{pN} \right> =
c^{\dagger}(z_1)...c^{\dagger}(z_{pN}) \left| 0 \right>  ,
\label{eq:electron_repr}
\\
\left| w_1...w_{hN} \right> =
c(w_1)...c(w_{hN}) \left| 1 \right>  ,
\label{eq:hole_repr}
\end{eqnarray}
where $\left| 0 \right>$ and $\left| 1 \right>$ respectively denote
the empty- and full-Landau level states, $c^{\dagger}(z)$ is
the operator creating an electron in a coherent state \cite{Girvin84}
centered at position $z$, $h$ is given by $h=q-p$
and where $pN$ and $hN$ are respectively the number
of electrons and holes in the lowest Landau level.

For $\beta=0$,
the density matrix of Eq. (\ref{eq:dens_mat_def}) is simply given by
\begin{eqnarray}
\rho_0 =
\prod_{j_1 <j_2}^{hN} ( w_{j_1} - w_{j_2} )
\prod_{i_1 <i_2}^{pN} ( z_{i_1} - z_{i_2} )
\nonumber \\ \times
\prod_{j=1}^{hN} \prod_{i=1}^{pN} ( z_i - w_j )
\prod_{j=1}^{hN} e^{-\frac{|w_j|^2}{4}}
\prod_{i=1}^{pN} e^{-\frac{|z_i|^2}{4}}
\label{eq:rho_beta=0}
\end{eqnarray}
in units of the magnetic length,
where $z_i$ and $w_j$ are respectively the complex coordinates
of electron $i$ and hole $j$.
Ignoring for a moment the distinction between variables
$z_i$ and $w_j$, we see that $\rho_{0}$ has
a functional form identical to that of the wave function describing
a full Landau level. 
Thus configurations $(w_1...w_{hN} z_1...z_{pN})$
characterized by non-negligeable values of $\rho_{0}$
have for feature ( denoted by (A) ) that their coordinates are
homogeneously distributed over the whole sample.
Let us now study the effect of the electron-electron repulsion
present in Eq. (\ref{eq:dens_mat_def}).
In the case of a repulsion characterized by vanishing
pseudo-potential coefficients \cite{Haldane87} for even values
of the electron-electron relative angular momentum,
the Hamiltonian matrix is given, up to an additive constant, by
\begin{eqnarray}
- \frac{\partial}{\partial \beta} \rho_{| \beta=0}
(w_1...w_{hN} z_1...z_{pN})
=
- \sum_{i=1}^{pN} \sum_{j=1}^{hN} \sum_{L}
\tilde{V}_L
\nonumber \\ \times
\int d^2 \xi 
\left[
{\cal P}_{
\psi_L^{(\xi)}(z_i w_j)
}
\rho_0
\right]
(w_1...w_{hN} z_1...z_{pN})  ,
\label{eq:Hamiltonian_matrix}
\end{eqnarray}
where the sum over $L$ is carried over odd positive integers,
$\tilde{V}_L$ denote the pseudo-potential coefficients,
$\psi_L^{(\xi)}(z_1 z_2)$ is the wave function describing two particles
with relative angular momentum $L$ and with center of mass in a coherent
state centered at position $\xi$ and where ${\cal P}$ is the projector
associated to this wave function, defined by
\begin{eqnarray}
\left[ {\cal P}_{ \psi_L^{(\xi)}(z_i w_j) } \rho_0 \right]
(w_1...w_{hN} z_1...z_{pN}) =
\psi_L^{(\xi)}(z_i w_j)
\nonumber \times \\
\int d^2 z_i' d^2 w_j' { \psi_L^{(\xi)} }^* (z_i' w_j')
\rho_0 (w_1..w_j'..w_{hN} z_1..z_i'..z_{pN})  .
\label{eq:proj_def}
\end{eqnarray}
As can be seen in Eq. (\ref{eq:Hamiltonian_matrix}),
the effect of the interaction at finite values of $\beta$
is to increase $\rho_{\beta}$ for configurations
which have for feature ( denoted by (B) ) that
their coordinates $w_j$ are close to their coordinates $z_i$.
Thus it is likely that the quantity $\rho_{\beta}$
for small but finite $\beta$ will be largest for configurations
$(w_1...w_{hN} z_1...z_{pN})$ in the 2D plane
which can be divided into compact clusters of equal sizes
containing each $p$ positions $z_i$ and $h$ positions $w_j$,
which possess both features (A) and (B).

Let us now construct a form
for the density matrix satisfying the latter requierement
for the special case of the filling fractions $\nu=p/(2p+1)$.
The question of the temperature range in which this form
is valid will be discussed latter.
We propose to use the function given by
$\kappa(w_1...w_{hN} z_1...z_{pN})={\cal A}_w {\cal A}_z 
\hat{\kappa}(w_1...w_{hN} z_1...z_{pN})$ and by
\begin{eqnarray}
\hat{\kappa} = \!
\int \!
\prod_{n=1}^N
\left[
d^2 \xi_n
{\cal P}_{
\chi^{(\xi_n)}( z_{p(n-1)+1}...z_{pn} w_{h(n-1)+1}...w_{hn} )
}
\right]
\rho_{0}  ,
\label{eq:dens_mat_form}
\end{eqnarray}
where ${\cal A}_w$ and ${\cal A}_z$ are antisymmetrization operators
acting respectively on sets of variables
$(w_1...w_{hN})$ and  $(z_1...z_{pN})$ and where $\chi^{(\xi_n)}$
is a function describing the correlation of $p$ electrons and $h=p+1$ holes
in a cluster centered at position $\xi_n$, given by
\begin{eqnarray}
\chi^{(\xi)}( z_1..z_p w_1..w_h) =
\prod_{i_1 < i_2}^p ( z_{i_1} - z_{i_2} )
\prod_{j_1 < j_2}^h ( w_{j_1} - w_{j_2} )^3
\nonumber \\ \times
\prod_{i=1}^{p} e^{\frac{\xi^* z_i}{2}-\frac{|z_i|^2+|\xi|^2}{4}}
\prod_{j=1}^{h} e^{\frac{\xi^* w_j}{2}-\frac{|w_j|^2+|\xi|^2}{4}}
\label{eq:Chi}  .
\end{eqnarray}
The projectors and the term $\rho_0$ in Eq. (\ref{eq:dens_mat_form})
respectively ensure that configurations leading to a large value
of $\kappa$ have features (B) and (A).
The integrals implicit in the projectors
of Eq. (\ref{eq:dens_mat_form}) can be evaluated using the identity
\begin{eqnarray}
\lefteqn{
\int \prod_{i=1}^n \left[
d^2 z_i e^{-\frac{|z_i|^2 + \xi z_i^*}{2}}
\right]
P_k^*(z_1..z_n) Q_l(z_1..z_n) f(z_1..z_n) =
} \nonumber \\ &
f(\xi..\xi) \int \prod_{i=1}^n \left[
d^2 z_i e^{-\frac{ |z_i|^2 }{2}}
\right]
P_k^*(z_1..z_n) Q_l(z_1..z_n)  ,
\label{eq:int_formula}
\end{eqnarray}
where $f$ is a polynomial and where $P_k$ and $Q_l$ are homogeneous 
polynomials of total degree $k$ and $l$ with $k \leq l$
and are invariant under global translations of their variables
\cite{foot1}.
This leads, up to a multiplicative constant, to
\begin{eqnarray}
\lefteqn{
\hat{\kappa} =
\int \prod_{n=1}^{N} \left[ d^2 \xi_n
\chi^{(\xi_n)}( z_{p(n-1)+1}...z_{pn} w_{h(n-1)+1}...w_{hn} )
\right]
} \nonumber \\ &
\times
\prod_{n_1<n_2}^N ( \xi_{n_1} - \xi_{n_2} )^{q^2}
\prod_{n=1}^N e^{-\frac{q|\xi_n|^2}{4}}  . \makebox[16mm]{}
\label{eq:dens_mat_simple}
\end{eqnarray}
The function $\chi$ is choosen so as to satisfy the following criteria:
(a) $\chi^{(\xi)}$ is antisymmetric under interchange of two
electron coordinates or under interchange of two hole coordinates.
(b) The total degree of the polynomial part of $\chi^{(0)}$ 
equals $q(q-1)/2$.
Indeed, in light of Eq. (\ref{eq:int_formula}), $\kappa$ would vanish
for a smaller degree.
On the other hand, a larger degree would lead \cite{foot1} to a reduction
of the power of factors $( \xi_{n_1} - \xi_{n_2} )$ in Eq.
(\ref{eq:dens_mat_simple}) and configurations with a large value
of $\kappa$ would not satisfy condition (A) anymore.
(c) Factors of type $(z_i - w_j)$ are absent so that $\chi$ is large
for configurations satisfying condition (B).

In order to support our claim that the function $\kappa$
provides a realistic description of the density matrix
for some temperature,
we now make the case that the groundstate wave functions derived
from this form for the density matrix are generalizations
of the pair wave functions of Ref. \cite{Morf86},
which are known to be accurate.
Let us denote by $\Phi(z_1...z_{pN})$ and $\Theta(w_1...w_{hN})$
the wave functions describing the groundstate at $\nu=p/(2p+1)$
in the electron- and hole-representations of Eqs.
(\ref{eq:electron_repr}) and (\ref{eq:hole_repr}), respectively.
If $\kappa$ provides a good description of the density matrix,
then $\Phi$ and $\Theta$ should maximize the quantity
\begin{equation}
R(\Phi,\Theta)=
\frac{
\int d^{2pN} Z d^{2hN} W \kappa^*(W,Z) \Phi(Z) \Theta(W)
}{
\sqrt{
\int d^{2pN} Z \left| \Phi(Z) \right|^2
\int d^{2hN} W \left| \Theta(W) \right|^2
} }  ,
\label{eq:GS_WF_criterium}
\end{equation}
where $Z \equiv (z_1...z_{pN})$ and $W \equiv (w_1...w_{hN})$.

We now approximate $R(\Phi,\Theta)$ by substituting $\hat{\kappa}$
for $\kappa$ in Eq. (\ref{eq:GS_WF_criterium}).
Ignoring for the moment the constraint of antisymmetrization,
we consider wave functions of the type
\begin{eqnarray}
\hat{\Phi} (z_1...z_{pN}) =
\prod_{i=1}^{pN} e^{-\frac{|z_i|^2}{4}}
\prod_{n=1}^N a_{l_a} (z_{(n-1)p+1}...z_{np})
\nonumber \\ \times
\prod_{n_1<n_2}^N
b_{l_b} (z_{(n_1-1)p+1}...z_{n_1p},z_{(n_2-1)p+1}...z_{n_2p})
\label{eq:el_ansatz}
\end{eqnarray}
where $a_{l_a}$ and $b_{l_b}$ are translationally invariant
and homogeneous polynomials of total degree $l_a$ and $l_b$ to be
determined.
The polynomial $a_{l_a}(z_{(n-1)p+1}...z_{np})$
describes correlations within the group $n$
of variables $z_{(n-1)p+1}...z_{np}$.
The polynomial 
$b_{l_b} (z_{(n_1-1)p+1}...z_{n_1p},z_{(n_2-1)p+1}...z_{n_2p})$
is built using products of $l_b$ factors of type
$(z_{(n_1-1)p+i_1} - z_{(n_2-1)p+i_2})$, with $i_1,i_2=1...p$,
and describes correlations between groups $n_1$ and $n_2$.
We write similarly
\begin{eqnarray}
\hat{\Theta} (w_1...w_{hN}) =
\prod_{j=1}^{hN} e^{-\frac{|w_j|^2}{4}}
\prod_{n=1}^N c_{l_c} (w_{(n-1)h+1}...w_{nh})
\nonumber \\ \times
\prod_{n_1<n_2}^N
d_{l_d} (w_{(n_1-1)h+1}...w_{n_1h},w_{(n_2-1)h+1}...w_{n_2h})  .
\label{eq:hole_ansatz}
\end{eqnarray}
We now introduce Eqs.
(\ref{eq:dens_mat_simple}),
(\ref{eq:el_ansatz}), and
(\ref{eq:hole_ansatz}) into Eq.
(\ref{eq:GS_WF_criterium})
and integrate over coordinates $z_i$ and $w_j$
before $\xi_n$ in the numerator.
In order to maximize the numerator, we demand that the result
of the integration over electron and hole coordinates
have the same dependence in variables $\xi_n$ as the kernel
of $\hat{\kappa}$ given by the second line of Eq. (\ref{eq:dens_mat_simple}).
Using Eq. (\ref{eq:int_formula}),
it can be seen by inspection \cite{foot1} that this is possible only when
$l_a$ and $l_c$ are equal to the degrees of polynomial parts of Eq.
(\ref{eq:Chi}) depending respectively on coordinates $z_i$ and $w_j$,
i.e. $l_a=p(p-1)/2$ and $l_c=3h(h-1)/2$, and when $l_b+l_d=q^2$.
The latter condition, together with the constraint that each coordinate
in $\hat{\Phi}$ or in $\hat{\Theta}$ have $qN-1$ zeros, leads to
$l_b=pq$ and $l_d=hq$.
In light of Eqs. (\ref{eq:int_formula}) and (\ref{eq:Chi}),
it is judicious to choose
\begin{eqnarray}
a_{l_a}(z_1...z_p)=\prod_{i_1 < i_2}^p ( z_{i_1} - z_{i_2} )
\label{eq:a}
\\
c_{l_c}(w_1...w_h)=\prod_{j_1 < j_2}^h ( w_{j_1} - w_{j_2} )^3
\label{eq:c}
\end{eqnarray}
in order to obtain a large result from integration over particle
and hole coordinates in the numerator
of Eq. (\ref{eq:GS_WF_criterium}).
Among polynomials $b_{pq}$ and $d_{hq}$ leading to the same
value for the numerator, those yielding to the most homogeneous
distribution of factors $(z_{i_1}-z_{i_2})$ over pairs
$1 \leq i_1 < i_2 \leq pN$
and of factors $(w_{j_1}-w_{j_2})$ over pairs $1 \leq j_1 < j_2 \leq hN$
lead to the smallest denominator.
We therefore choose
\begin{eqnarray}
b_{pq} (z_1...z_{2p}) =
\prod_{i_1=1}^p \prod_{i_2=1}^p
( z_{i_1} - z_{p+i_2} )^2
\nonumber \\ \times
\sum_{P_p} \prod_{i=1}^p
( z_{P_i} - z_{p+i} )
\label{eq:b}
\end{eqnarray}
and
\begin{eqnarray}
d_{hq} (w_1...w_{2h}) =
\prod_{j_1=1}^h \prod_{j_2=1}^h
( w_{j_1} - w_{h+j_2} )^2
\nonumber \\ \times
\sum_{P_h} \prod_{j=1}^h
( w_{P_j} - w_{h+j} )^{-1}
\label{eq:d}
\end{eqnarray}
where $P_p$ and $P_h$ respectively denote sums over permutations
of $p$ and $h$ objects.

The wave functions $\Phi={\cal A}_z \hat{\Phi}$
and $\Theta={\cal A}_w \hat{\Theta}$ defined using Eqs.
(\ref{eq:el_ansatz}), (\ref{eq:hole_ansatz}),
(\ref{eq:a}), (\ref{eq:c}), (\ref{eq:b}) and (\ref{eq:d})
are identical to the pair wave functions of Ref. \cite{Morf86}
when $p=2$ and $h=2$, respectively.
$\Phi$ and $\Theta$ provide a good microscopic
description of groundstates at filling factors $\nu=p/(2p+1)$
and $\nu=h/(2h-1)$, respectively. Indeed, they lead
to numbers of flux quanta which are in agreement with the prescription
of the hierarchical scheme \cite{Haldane87} and have a large overlap with
the exact groundstates of the corresponding systems of particles
on a sphere in presence of Coulomb repulsion: In the case $\nu=2/3$
and for 8 and 10 particles respectively, the overlaps are $0.954$
and $0.930$ whereas the numbers of $(L=0)$ states are 2 and 6.
In the case $\nu=3/5$ and for 6 and 9 particles respectively,
the overlaps are $0.988$ and $0.970$ and the numbers
of $(L=0)$ states are 3 and 8.

We now study the constraints imposed by statistics on allowed values
of parameters $p$, $h$ and $q=p+h$ using a path integral
representation of the partition function
\begin{equation}
{\cal Z}(\beta)=
\int d^2 z_1...d^2 z_{pN}
\left<z_1...z_{pN}| e^{ - \beta \tilde{V} } |z_1...z_{pN}\right>  .
\label{eq:partition_funct}
\end{equation}
The exponential in Eq. (\ref{eq:partition_funct}) is brocken into $M$
pieces $\exp( - \beta \tilde{V}/M )$ with $M$ even,
and projectors $|z_1^{(m)}...z_{pN}^{(m)}><z_1^{(m)}...z_{pN}^{(m)}|$
and $|w_1^{(m)}...w_{hN}^{(m)}><w_1^{(m)}...w_{hN}^{(m)}|$ are inserted
alternatively at imaginary times $m\beta/M$ for $m$ even
and $m$ odd, respectively.
We consider configurations which can be divided at all imaginary times
in clusters containing $p$ coordinates $z_i^{(m)}$ and $h$ coordinates
$w_j^{(m)}$, which significantly contribute to the partition function.
For simplicity we further demand that $z_i^{(m)}$ and $w_j^{(m)}$ have
a smooth dependence on $m$ and belong to the same cluster throughout
imagimary time.
The phase corresponding to a given path is a product of phase contributions
associated to each time step, which can in turn be evaluated by means
of Eq. (\ref{eq:rho_beta=0}).
Ignoring factors $( z_i - w_j )$ for $z_i$ and $w_j$ belonging to the
same cluster,
this leads to $(-1)^{ph[P]}$, where $[P]$ is the parity of the permutation
taking place among clusters between imaginary time $\tau=0$ and $\tau=\beta$.
Thus such paths contribute constructively to the partition function only if
$ph$ is even, that is if (a) $p$ is even and $h$ is odd
or (b) $p$ is odd and $h$ is even or (c) both $p$ and $h$ are even.
Note that cases (a) and (b) imply a filling factor $\nu=p/q$ with
odd denominator.

The fact that all incompressible fractional quantum Hall states
observed in the lowest Landau level are characterized by a filling factor
with odd denominator may be attributed to the fact that a stable cluster
made up of $p$ electron coordinates and $h$ hole coordinates can only be
obtained
if one of the parameters is odd, thus corresponding to case (a) or (b).
Indeed, for $p$ and $q$ both even, the paths obtained by clustering
$p$ coordinates $z_i$ and $h$ coordinates $w_j$ could be energetically
unfavorable compared with paths obtained by clustering
$p/2$ coordinates $z_i$ and $h/2$ coordinates $w_j$.

The incompressible state at $\nu=5/2$ is characterized by $p=h=2$
( case (c) ).
The absence of incompressible state at $\nu=1/2$ is attributed
\cite{Ho95} to the fact that, in the lowest Landau level, clusters
with $p=h=2$ are unstable toward the formation of clusters
with $p=h=1$. 
Since the paths of such $p=h=1$ clusters contribute to the partition
function with a phase $(-1)^{[P]}$ reminiscent of Fermions,
they can be regarded as composite Fermions \cite{Halperin93}.

It is important to note that the approximation to the density
matrix provided by function $\kappa$ of Eq. (\ref{eq:dens_mat_simple})
is unable to describe neutral excitations consisting in pairs
of well-separated fractionally-charged excitations.
Indeed, when using this approximate form of the density matrix
to evaluate the number of electrons in a surface large compared
to the squared magnetic length, one obtains fluctuations much too
small to include the effect of such excitations.
In order to include the effect of a single quasiparticle-quasihole
pair, it is necessary to add to $\kappa$ a further term. 
This additional term can be obtained from the right-hand side of Eq.
(\ref{eq:dens_mat_form}) by replacing one of the projectors associated
to clusters of $p$ electrons and $h$ holes by two projectors:
One associated to a cluster of $p'$ electrons and $h'$ holes
and an other associated to a cluster of $p-p'$ electrons and $h-h'$
holes, with $p'$ and $h'$ satisfying $p'h-ph'= \pm 1$.
The cluster associated to the plus sign leads to a local charge defect
$e/q$ (quasiparticle) and the other to $-e/q$ (quasihole),
where $e$ is the electron's charge.
The parameters $p'$ and $h'$ are given by
$p'/h'=1/(n_1+1/(n_2+...+1/n_{k-1}))...))$ in terms of the continued
fraction decomposition
$p/h=1/(n_1+1/(n_2+...+1/n_k))...))$.
Thus the creation of a neutral excitation corresponds to the breaking
of a cluster of $p$ electrons and $h$ holes.
The fact that this breaking requires a finite amount of potential energy,
together with the fact that $\kappa$ of Eq. (\ref{eq:dens_mat_form})
considered as a function of $z_1...z_{pN}$ or of $w_1...w_{hN}$
has a small overlap with the contribution described above
including a neutral excitation,
leads to a gap in the energy spectrum and to incompressibility.

We finally address the question of the temperature range in which
the function $\kappa(w_1...w_{hN} z_1...z_{pN})$
of Eq .(\ref{eq:dens_mat_form}) may be adequate
to describe the density matrix of Eq. (\ref{eq:dens_mat_def}).
Although $\kappa$ is unable to describe neutral excitations consisting
in pairs of well-separated elementary excitations
which are associated to short wavelength collective modes \cite{Laughlin87},
We do believe that the effect of long wavelength collective modes
are included in $\kappa(w_1...w_{hN} z_1...z_{pN})$ and are responsible
for the correlations existing between variables $z_i$ and $w_j$.
This collective modes are characterized by a finite excitation energy
$\epsilon_{k=0}$.
We thus conjecture that the temperature for which $\kappa$ is adequate
to describe the density matrix is of the order of $\epsilon_{k=0}/k_B$.

\section{Acknowledgements}

We would like to thank R. Morf for helpfull discussions.
We acknowledge support from the Swiss National Science Fundation
under grand No. 2000-040395.94/1.


\begin{references}
\bibitem[9]{foot1}
This identity can be established using a transformation to
center of mass and relative coordinates.
If $k>l$, the left-hand side of Eq. (\ref{eq:int_formula})
is proportional to a linear combination of terms of the type
$\partial_{z_1}^{l_1} ... \partial_{z_n}^{l_n} f_{|z_1=...=z_n=\xi}$
with $l_1+...+l_n=k-l$.
\end{references}
\end{document}